\documentclass[preprint,showpacs,preprintnumbers,amsmath,amssymb]{revtex4}
\usepackage{epsfig}
\usepackage{graphicx}
\usepackage{dcolumn}
\usepackage{bm}

\let\jnfont=\rm
\def\NPB#1,{{\jnfont Nucl.\ Phys.\ }{\bf B#1},}
\def\PLB#1,{{\jnfont Phys.\ Lett.\ B }{\bf #1},}
\def\PRD#1,{{\jnfont Phys.\ Rev.\ D }{\bf #1},}
\def\PRL#1,{{\jnfont Phys.\ Rev.\ Lett.\ }{\bf #1},}
\def\ZPC#1,{{\jnfont Z.\ Phys.\ C }{\bf #1},}
\def\EPJC#1,{{\jnfont Eur.\ Phys.\ J.\ C }{\bf #1},}

\def\lsim{\mathrel{\mathpalette\oversim<}}

\def\oversim#1#2{\lower0.5ex\vbox{\baselineskip0pt\lineskip0pt
  \lineskiplimit0pt\everycr{}\tabskip0pt
  \halign{$\mathsurround0pt #1\hfil##\hfil$\crcr #2\crcr\sim\crcr}}}

\begin{document}
\draft
\preprint{}

\title{Lepton-Specific Two-Higgs Doublet Model: Experimental Constraints and
       Implication on Higgs Phenomenology}

 \author{Junjie Cao$^1$,  Peihua Wan$^1$, Lei Wu$^1$, Jin Min Yang$^{2,3}$}

\address{
 $^1$ College of Physics $\&$ Information Engineering,
      Henan Normal University, Xinxiang 453007, China \\
 $^2$ Key Laboratory of Frontiers in Theoretical Physics,
      Institute of Theoretical Physics, Academia Sinica, Beijing 100190, China\\
$^3$ Kavli Institute for Theoretical Physics China, Academia Sinica, Beijing 100190, China
\vspace*{1cm}}

\begin{abstract}
We examine various direct and indirect constraints on the
lepton-specific two-Higgs doublet model and scrutinize the property
of the Higgs bosons in the allowed parameter space. These
constraints come from the precision electroweak data, the direct
search for Higgs boson, the muon anomalous magnetic moment, as well
as some theoretical consistency requirements. We find that in the
allowed parameter space the CP-odd Higgs boson $A$ is rather light
($m_A < 30$ GeV with $95\%$ possibility), which is composed dominantly
by the leptonic Higgs and decays dominantly into $\tau^+ \tau^-$;
while the SM-like Higgs boson $h$ (responsible largely for electroweak
symmetry breaking) decays dominantly in the mode $h \to A A \to 4
\tau $ with a large decay width,
which will make the Higgs discovery more difficult at the LHC.
Whereas, this scenario predicts a branching ratio $Br(Z \to \tau^+ \tau^-
A)$ ranging from $10^{-5}$ to $10^{-4}$, which may be accessible at the
the GigaZ option of the ILC.
\end{abstract}

\maketitle

\section{Introduction}
The phenomenological success of the standard model (SM) has
significantly limited the possibility of new physics except for
the Higgs sector which remains untested. There are numerous
speculations on the possible extensions of the Higgs sector, among
which the simplest is to introduce one more Higgs doublet.
Compared with the SM, such simple two-Higgs doublet models usually
have much more complicated Higgs phenomenology. In the SM  a
single Higgs doublet is responsible for the electroweak symmetry
breaking and the Higgs couplings with fermions and gauge bosons
are completely determined by their masses, and therefore there is
little guesswork in determining the discovery channels for the
Higgs boson \cite{SM-Higgs}. In the two-Higgs doublet models,
however, the addition of new  scalars and the modification of the
Higgs interactions will significantly complicate the Higgs
discovery at the LHC \cite{Ellwanger}. Given the imminent running
of the LHC, the phenomenological study of various such models is
urgently important.

In this paper we focus on a special two-Higgs doublet model
called the lepton-specific two-Higgs doublet model (L2HDM) \cite{L2HDM}.
Since this model is arguably well motivated
from some fundamental theory and also has some phenomenological virtues
(e.g., it can provide a natural explanation for the leptonic
cosmic ray signals reported by PAMELA and ATIC \cite{Dark matter}),
it has attracted much attention \cite{L2HDM-phenome1,L2HDM-phenome2}.
We will check various constraints on the model parameters and
then scrutinize the property of the Higgs bosons in the allowed
parameter space.
These constraints come from the precision electroweak data,
the direct search for Higgs boson, the muon anomalous magnetic moment,
as well as some theoretical consistency requirements.
 Our main observation is that in the allowed parameter space
the CP-odd Higgs boson $A$  must be light ($m_A <$ 30 GeV with $95\%$
possibility ), which is composed dominantly by the leptonic Higgs and
decays dominantly into $\tau \bar{\tau}$; while the SM-like Higgs
boson $h$ (responsible largely for electroweak symmetry breaking)
decays dominantly in the mode $h \to A A \to 4 \tau $ with a decay
width usually exceeding several tens of GeV, which may make the Higgs
discovery more difficult at the LHC.

This paper is organized as follows. In Sec. II we
recapitulate the L2HDM model.  In Sec. III we examine
various constraints on the parameter space and
study the properties of the Higgs bosons in the allowed
parameter space.
Finally, In Sec. IV we give our conclusion.

\section{The lepton-specific two-Higgs doublet model}
\label{sec2}
The L2HDM is a special  two-Higgs doublet model in which
one Higgs doublet  $\phi_1$ couples only to leptons while the other
doublet $\phi_2$ couples only to quarks. Both Higgs doublets
contribute to the electroweak symmetry breaking:
$v^2 = v_1^2 + v_2^2 = ( 246 ~{\rm GeV})^2 $ with $v_1$ and $v_2$
being respectively the vacuum expectation values of $\phi_1$
and $\phi_2$; whereas their relative contributions can be
quite different and can be parameterized by the ratio
$\tan \beta = v_2/v_1$. So for a large $\tan \beta$ the
lepton Yukawa couplings can be greatly enhanced.

The Yukawa interactions and the Higgs potential are
given by \cite{L2HDM-phenome2}
\begin{eqnarray}
\mathcal{L}_{\mathit{Y}}&=& -Y_e^{ij}\bar{\ell}_i\phi_1 e_j
     - Y_u^{ij} \bar{q}_i\phi_2^c u_j - Y_d^{ij} \bar{q}_i\phi_2 d_j+h.c. \\
V &=&  m_{1}^2|\phi_1|^2+m_{2}^2|\phi_2|^2
  - \left(m_{3}^2\phi_1^{\dagger}\phi_2+h.c.\right)
  +\frac{\lambda_{1}}{2}|\phi_1|^4+
  \frac{\lambda_{2}}{2}|\phi_2|^4\nonumber \\
&&
  +\lambda_{3}|\phi_1|^2|\phi_2|^2 +\lambda_{4}|\phi_1^{\dagger}\phi_2|^2  +
  \frac{\lambda_5}{2}\left[(\phi_1^\dagger \phi_2)^2+h.c.\right],
\label{eq:Vstandard}
\end{eqnarray}
where $i,j$ are generation indices, $Y_e$, $Y_u$ and $Y_d$
are $3\times 3$ Yukawa matrices, $q_i$ and $\ell_i$  denote respectively the
the left-handed quark and lepton fields, $u_i$ and $d_i$ denote respectively
the right-handed up- and down-type quark fields, $e_i$ denotes the right-handed
lepton fields, and $m^2$ and $\lambda$ are free parameters.

Just like the usual two-Higgs doublet model \cite{Haber},
the spectrum of the Higgs sector includes
three massless Goldstone modes, which become the longitudinal
modes of $W^\pm$ and $Z$ bosons, and five massive
physical states: two $CP$-even states $h$ and $H$,
a pseudoscalar $A$, and a pair of charged states $H^\pm$.
These states are related to the doublets $\phi_1$ and $\phi_2$ by
\begin{eqnarray}
\phi_1^0 & =& \frac{1}{\sqrt{2}} ( v_1 + H\cos\alpha  - h\sin\alpha
 + i  G^0 \cos\beta - i A \sin\beta  ) , \\
\phi_2^0 & =& \frac{1}{\sqrt{2}} ( v_2 + H \sin\alpha  + h\cos\alpha
 + i G^0 \sin\beta  + i A\cos\beta  ) , \\
\phi_1^\pm & =& G^\pm \cos\beta - H^{\pm} \sin\beta  , \\
\phi_2^\pm & =& G^\pm \sin\beta  + H^{\pm}\cos\beta  ,
\label{translation}
\end{eqnarray}
where $ \alpha $ is the mixing angle that diagonalizes the
mass matrix of the CP-even Higgs fields.

Due to the constraint $v_1^2+v_2^2= (246\ {\rm GeV})^2$,  the eight
free parameters in Eq.~(\ref{eq:Vstandard}), i.e., $\lambda_i$ $(i=1,\ldots,5)$
and $m^2_i$ (i=1,2,3), reduce to seven. In our analysis we choose
the following seven parameters as the input parameters of the L2HDM:
\begin{equation}
m_h, ~m_H, ~m_A, ~m_{H^\pm}, ~\tan\beta, ~\sin\alpha, ~\lambda_{5},
\label{eq:MeaningfulBasis}
\end{equation}
where $m_h$, $m_A$, $m_H$, and $m_{H^\pm}$ are the masses of the
corresponding physical states. Throughout this paper, we use $H$
($h$) to denote the Higgs boson with $\phi_1^0$ ($\phi_2^0$) as its
dominant component, which means that we choose $\cos^2 \alpha >
1/2$.

The interactions of the
Higgs physical states with fermions are then given by \cite{L2HDM-phenome2}
\begin{eqnarray}
{\mathcal{L}_{\mathit{Y}}}&=&
  - \frac{g m_{e_i}}{2 m_W \cos\beta} ( \cos\alpha ~\bar{e_i} e_i H
  - \sin\alpha ~\bar{e_i} e_i h ) \nonumber \\
  && - \frac{g m_{q_i}}{2 m_W \sin\beta} ( \sin\alpha ~\bar{q_i} q_i H
 + \cos\alpha ~\bar{q_i} q_i h )  \nonumber \\
  &&
+ \frac{i g m_{u_i}}{2 m_W}  \cot \beta ~ \bar{u_i} \gamma_5 u_i A
- \frac{i g m_{d_i}}{2 m_W}  \cot \beta  ~\bar{d_i} \gamma_5
  d_i A \nonumber \\
&& + \frac{i g m_{e_i} }{2 m_W}\tan\beta ~\bar{e_i} \gamma_5 e_i A \nonumber \\
&& + \frac{g  V_{ij} }{\sqrt{2} m_W}\cot\beta ~\bar{u}_i ( m_{u_i} P_L
   - m_{d_j} P_R ) d_j  H^+ \nonumber \\
 && + \frac{g m_{e_i}}{\sqrt{2} m_W} \tan\beta~\bar{\nu}_i P_R e_i   H^+
\label{couplings}.
\end{eqnarray}
Obviously, for a large $\tan\beta$ the lepton Yukawa couplings are greatly enhanced
relative to the SM prediction. One can also check that the
couplings of $Z Z h$ and $Z Z H$  are given by
\begin{eqnarray}
V_{ZZh} &=& \frac{g m_Z}{\cos \theta_W} \sin (\beta - \alpha ), \\
V_{ZZH} &=& \frac{g m_Z}{\cos \theta_W} \cos (\beta - \alpha ),
\end{eqnarray}
which satisfy the sum rule $V_{ZZh}^2 + V_{ZZH}^2 = V_{ZZh_{SM}}^2$.
For  a large $\tan\beta$ (this is the case required by the
experimental constraints, as shown below), the coupling $Z Z h$ is
dominant over $Z Z H$, so $h$ is usually called the SM-like Higgs
boson.

\section{Constraints on the L2HDM}
\label{sec3}
We note that both the
theoretical consistency and the electroweak data have
limited the parameter space of the L2HDM.
In our study we consider the following theoretical
constraints:
\begin{itemize}
\item[(1)] The perturbativity is valid in the Higgs sector, which
requires $\lambda_i < 4\pi$ ($i=1,\ldots,5$). \item[(2)] The
$S$-matrix satisfies all relevant tree-unitarity constraints,
which implies that the quartic couplings $\lambda_i$ satisfy
\cite{Unitary}
\begin{eqnarray}
&&
3(\lambda_1+\lambda_2)\pm\sqrt{9(\lambda_1-\lambda_2)^2+4(2\lambda_3+\lambda_4|)^2}
<16 \pi,  \nonumber \\
&&
\lambda_1+\lambda_2\pm\sqrt{(\lambda_1-\lambda_2)^2+4|\lambda_5|^2}
<16\pi,  \nonumber \\
&&
\lambda_1+\lambda_2\pm\sqrt{(\lambda_1-\lambda_2)^2+4|\lambda_5|^2}
<16 \pi,  \nonumber \\
&&\lambda_3+2\lambda_4\pm 3 | \lambda_5|<8\pi, \nonumber\\
&& \lambda_3\pm\lambda_4 < 8\pi,\nonumber\\
&& \lambda_3\pm|\lambda_5|<8\pi.
\label{eq:PertBounds}
\end{eqnarray}
\item[(3)] The scalar potential in Eq.~(\ref{eq:Vstandard}) is finite at large
field values and contains no flat directions, which translate
into the bounds \cite{Haber}
\begin{eqnarray}
&&  \lambda_{1,2}>0,  \nonumber \\
&&  \lambda_3>- \sqrt{\lambda_1\lambda_2}, \nonumber \\
&&  \lambda_3+\lambda_4-|\lambda_5|> - \sqrt{\lambda_1\lambda_2}.
\label{eq:VacStabBounds}
\end{eqnarray}
\end{itemize}
On the experimental side, we consider the following constraints:
\begin{itemize}

\item[(4)] The lower mass bound on the charged Higgs bosons: $m_{H^+} >
92 $ GeV \cite{OPALchargedHiggs}.

\item[(5)] The constraints from the LEP search for neutral Higgs
bosons. We compute the signals from the Higgsstrahlung production
$e^+ e^- \to Z H_i $ ($H_i= h, H$) with $H_i \to 2 b, 2 \tau, 4 b, 4
\tau, 2 b 2 \tau $ \cite{LHWGSM,Opal3} and from the associated
production $e^+ e^- \to H_i A$ with $H_i A \to 4 b, 4 \tau, 2 b 2
\tau, 6 b, 6 \tau$ \cite{DELPHI} and compare them with their LEP
data. We also consider the constraints from $e^+ e^- \to Z H_i $ by
looking for a peak of $M_{H_i}$ recoil mass distribution of
$Z$-boson \cite{ALEPH} and the constraint of
$\Gamma ( Z \to H_i A) < 5.8$ MeV when $m_A + m_{H_i} < m_Z $ \cite{Monig}.

\item[(6)] The constraints from the LEP search for a light Higgs boson
via the Yukawa process $e^+ e^- \to f \bar{f} S$ with $f=b, \tau$
and $S$ denoting a scalar \cite{ffS}. These constraints can limit
$f\bar{f}S$ coupling versus $m_S$ and thus can constrain the parameters
of the L2HDM.

\item[(7)] The constraints from the $W$-boson mass.
The L2HDM Higgs sector can shift the $W$-boson mass through radiative corrections.
We require the corrected $W$-boson mass to lie within the $2 \sigma$ range
of the global-fit value. The SM prediction for the $W$-boson mass is 80.363
GeV for $m_t = 173$ GeV and $m_H = 111 $ GeV \cite{LEP-Report}, and
its fitted value is $ 80.398 \pm 0.025 $ GeV \cite{PDG}. We use the
formula in \cite{Cao} in calculating the mass and consider
the effect of a different top quark mass (in our calculation we take $m_t =
171.3 $ GeV). We also subtract the contribution from the SM Higgs
boson to avoid double counting the contribution from the Higgs
sector.

\item[(8)] The constraints from $Z \tau^+ \tau^-$ coupling.
For a large $\tan \beta$ the L2HDM Higgs sector can give
sizable radiative corrections to $Z \tau^+ \tau^-$ coupling.
We calculate such corrections and require the corrected $Z \tau^+ \tau^-$
coupling to lie within the $2 \sigma $ range of its fitted
value. The SM prediction for this coupling at $Z$-pole is given by
$g_V^{SM}=-0.03712 $ and $g_A^{SM} = -0.50127 $ \cite{LEP-Report}, and
the fitted value given respectively by $ -0.0366 \pm 0.00245 $ and
$ -0.50204 \pm 0.00064 $\cite{LEP-Report}. We use the formula in \cite{Cao} in
our calculation.

\item[(9)]  The constraints from $\tau$ leptonic decay.
We require the  L2HDM correction to the branching
ratio $Br( \tau \to e \bar{\nu}_e \nu_\tau)$ to be
in the range of $-0.80\% \sim 1.21\%$ \cite{tau-decay}.
We use the formula in \cite{tau-decay} in our calculation.

\item[(10)]  The constraints from the muon anomalous magnetic moment $a_\mu$.
Now both the theoretical prediction and the experimental measured
value of $a_\mu$ have reached a remarkable precision, but a
significant deviation still exists: $a_\mu^{exp} - a_\mu^{SM} = (29
\pm 8.8 ) \times 10^{-10}$ \cite{mug-2}. In our analysis we require
the L2HDM to account for such difference at $2 \sigma $ level. Note
that in the L2HDM, $a_\mu$ gets additional contributions from the
one-loop diagrams induced by the Higgs bosons and also from the
two-loop Barr-Zee diagrams mediated by $A$, $h$ and
$H$ \cite{Barr-Zee}. If the Higgs bosons are not too light,
the contributions
from the Barr-Zee diagrams are more important. To account for the
discrepancy of $a_\mu$, one needs a light $A$ along with a
large $\tan \beta$ to enhance the effects of the Barr-Zee
diagram involving the $\tau$-loop. The CP-even Higgs bosons
 are usually preferred to be heavy since their contribution to
$a_\mu$ is negative.

\item[(11)] Since the CP-odd Higgs $A$ can be quite light and
$h,H \to A A$ may open up with a large decay width,
we require the width of any Higgs boson in the
L2HDM to be smaller than its mass (otherwise the Higgs boson
may be too fat).
\end{itemize}
With the above constraints, we scan the parameter space of the L2HDM
in the ranges:
\begin{eqnarray}
&& 1 < \tan\beta < 200, \quad  5 {\rm ~GeV} < m_A <100 {\rm ~GeV}, \nonumber \\
&& 5 {\rm ~GeV} <m_{h,H} < 350 {\rm ~GeV}, \quad 92 {\rm ~GeV} <
m_{H^+} < 350 {\rm ~GeV}, \nonumber \\
&&-\sqrt{2}/2 < \sin\alpha < \sqrt{2}/2, \quad |\lambda_5| < 4 \pi.
\end{eqnarray}
\begin{figure}[htbp]
\epsfig{file=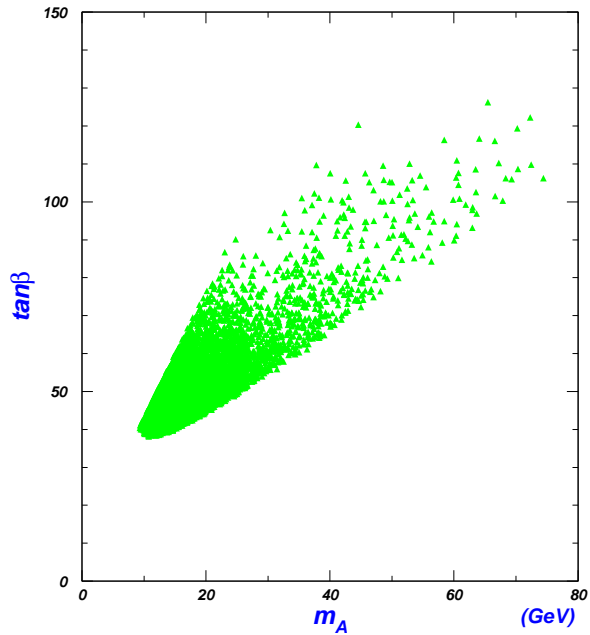,width=8cm} \vspace*{-0.6cm} \caption{Scatter
plots of the allowed parameter space in the plane of $\tan\beta$ versus $m_A$.}
\end{figure}

With $10^{12}$ random samplings, we get the allowed parameter space
shown in Figs.1-2. Fig.1 shows that the allowed  parameter space has
a light $A$ ( $ m_A \lsim 80$ GeV) and a large  $\tan \beta$ ( $37
\lsim \tan \beta \lsim 130$), which mainly comes from the
explanation of the $a_\mu$ discrepancy.
Among the surviving samples displayed in Fig.1, about  $95 \%$
satisfy $m_A < 30$ GeV and about $70\%$ satisfy $m_A < 20$ GeV, which
means that a very light $A$ is highly preferred by the constraints.

\begin{figure}[tb]
\epsfig{file=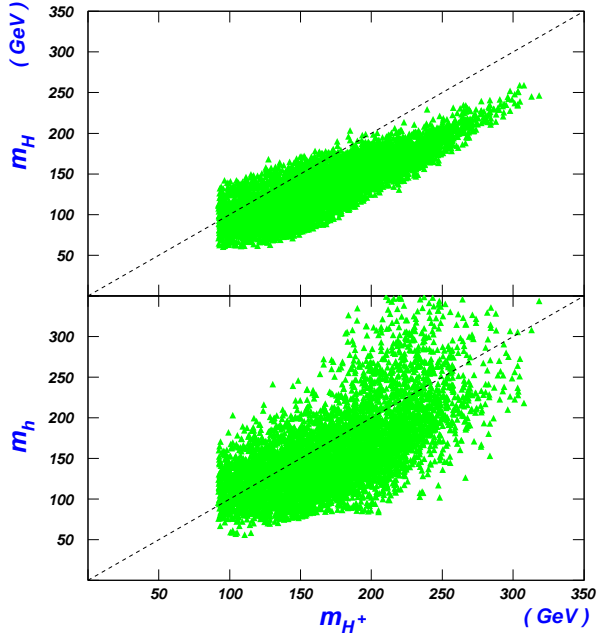,width=8cm} \vspace*{-0.4cm} \caption{Same as
Fig.1, but projected in the planes of $m_h$ and $m_H$ versus $m_{H^+}$.}
\end{figure}

Fig.2 shows the allowed parameter space projected in the planes of
$m_h$ and $m_H$ versus $m_{H^+}$.  Three characters should be noted
about this figure. The first is that all the Higgs bosons are
lighter than 350 GeV (lighter than 250 GeV for about $90\%$ of the
surviving samples), which is mainly due to the unitary requirement
and the $a_\mu$ constraint. The second is that $h$ and $H$ can be as
light as 58 GeV because the LEP2 bound is relaxed significantly due
to the weakened $Z Z h$ and $Z A H$  couplings by the seizable
mixing angle $\alpha$  and the open-up of the new decay mode $H,h
\to A A \to 4 \tau$ \footnote{For $m_A < 10 $ GeV, the constraint
from LEP search for Higgs bosons is still strong and in this case,
the lower mass bound of $h$ is 86 GeV \cite{2-mu-2-tau}.}. The third
character is that the values of both $m_h$ and $m_H$ are close to
the value of $m_{H^+}$, which is helpful in reducing the L2HDM
contribution to the precise electroweak data such as $m_W$ and $Z
\bar{\tau} \tau$ couplings at Z-pole. For $m_{H^+}
> 250$ GeV, the data require $|\sin (\beta - \alpha)| \sim 1$ \cite{Yuan}, and
in this case, $m_h$ has little effects on the data so that it can
deviate significantly from $m_{H^+}$.

In summary, the above results indicate that the preferred parameter
space of the L2HDM is $37 \lsim \tan \beta \lsim 80$, $m_A \lsim 30$ GeV
and the other Higgs bosons lighter than 250 GeV. Note that the
above favored region is obtained by considering all the constraints
(1-11), instead of any individual constraint. For example,
for $ \tan \beta > 200$, our results indicate that the CP-odd Higgs boson
$A$ as heavy as $120$ GeV can still explain $a_\mu$; but such a large
$\tan \beta$ is disfavored by the $Z \bar{\tau} \tau$ coupling at $Z$-pole
or by $\tau$ leptonic decay. Another point we should address is that in the
L2HDM, the processes $ B \to X_s \gamma $ and $\Upsilon \to A
\gamma$ cannot impose any further constraints \cite{L2HDM-phenome2}.
The reason is that in the surviving parameter space, $\tan \beta$
must be larger than 37 and, consequently, the couplings of bottom
quark with $H^+$ and $h$ are suppressed, as shown in
Eq.(\ref{couplings}). Finally, we would like to emphasize that in contrast
to the L2HDM which has a large parameter space to account
for the $a_\mu$ discrepancy without conflicting with other
experimental data, the popular type-II 2HDM is very difficult
to do so \cite{Barr-Zee}. This is one of the virtues of the L2HDM.

\section{Implication on Higgs phenomenology}
\label{sec4}
Eq.(\ref{couplings}) indicates that the lepton
couplings of $A$, $H$ and $H^+$ are enhanced by large $\tan \beta$,
while quark couplings are suppressed. Since the allowed parameter
space has a large $\tan \beta$,  the couplings of  $\tau$ lepton
with $A$, $H$ and $H^+$ are larger than the top quark couplings. So
these scalars will decay dominantly into $\tau$ leptons rather than
into top quarks (if kinetically allowed). Moreover, a light $A$ can
change the phenomenology of other Higgs bosons by opening new decay
modes like $h,H \to A A$, $h,H \to A Z $ and $H^+ \to A W^+$. As
discussed earlier, in case of a large $\tan \beta$ and a small
$\alpha$, $h$ is the SM-like Higgs boson, mainly responsible for the
electroweak symmetry breaking and couples to weak gauge bosons like
the SM Higgs. Therefore, the phenomenology of $h$ is of primary
importance and will be studied in the following.

\begin{figure}[htbp]
\epsfig{file=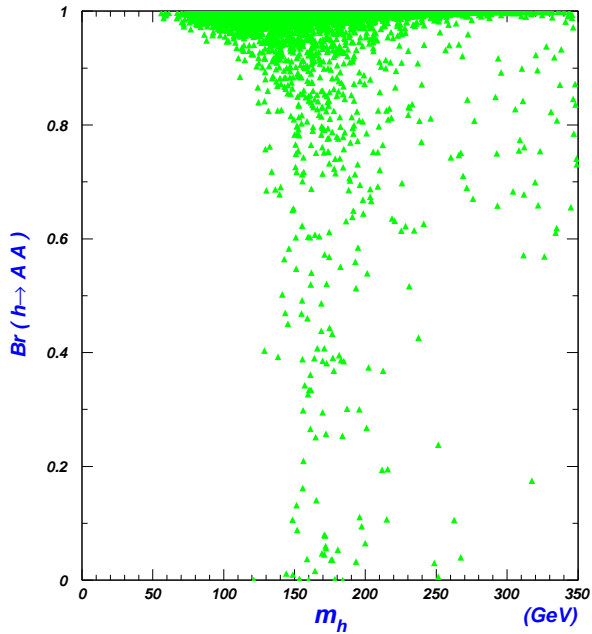,width=8cm} \vspace*{-0.5cm} \caption{Same as
Fig.1, but showing the branching ratio of $ h \to A A $ versus
$m_h$. }
\end{figure}

Fig.3 shows  the branching ratio of $h \to A A$ versus $m_h$. Here
we have considered all the decay modes of $h$ including $h \to V V,
A Z, \tau \bar{\tau}, b \bar{b}, t \bar{t}$. This figure shows that
for most of the allowed parameter space (about $99\%$), $h \to A A
\to \tau \bar{\tau} \tau \bar{\tau}$ is the dominant decay mode.
This will make the detection of $h$ difficult at the LHC because the
lightness of $A$ (note $m_A< 20$ GeV for about $70\%$ surviving
samples) will make the $\tau$ leptons from its decay highly
collimated \cite{Ellwanger,Han}, which is usually regarded as
a difficult scenario in Higgs discovery at the LHC in the next-to-minimal
supersymmetric model \cite{Ellwanger}. Another reason for the
detection difficulty of $h$ is that for more than $80\%$ of the
allowed parameter space, the width of $h$ is found to be larger than
10 GeV. Such a wide width will smear the peak of the invariant mass
distribution of $h$-decay products and make the detection more
difficult.

We note that in the L2HDM, $A\to \mu^+ \mu^-$ is the second largest
decay mode of $A$. So $h \to AA$ can give the multi-muon signal,
like the scenario proposed in \cite{hooper}. Unfortunately, in the
L2HDM the branching ratio of $A \to \mu^+ \mu^-$ is of order
$10^{-3}$, which will make the channel $h\to AA \to 4\mu$ quite
hopeless at the LHC. Note that some authors have considered the
channels $h\to AA \to 2\mu +2\tau$ \cite{2-mu-2-tau} and
$h\to AA \to 4\tau \to 2\mu+2~jets$ \cite{2-mu-2j} as well as the
diffractive Higgs production $pp\to pp+h$ followed by $h\to
4\tau$ \cite{diffrac} to detect such a $h$, but all these studies
did not consider the worse case of a fat $h$.
We also checked that the branching
ratio of $ h \to \gamma \gamma$ is usually suppressed to be less
than $10^{-6}$ and thus too small for the detection.

\begin{figure}[htbp]
\epsfig{file=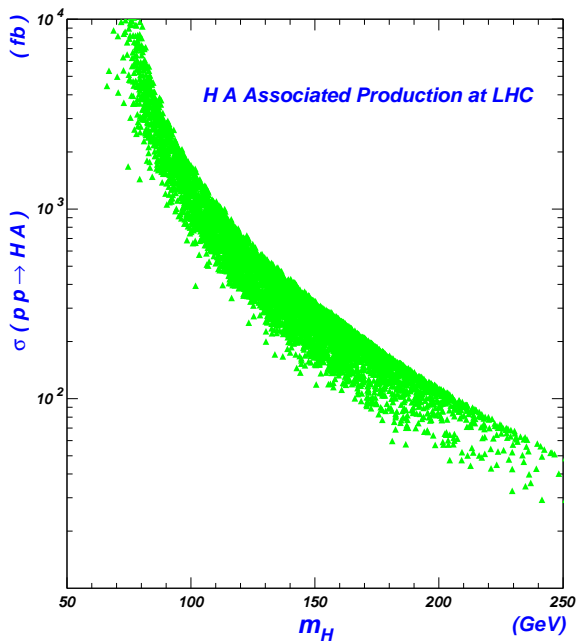,width=8cm} \vspace*{-0.5cm} \caption{Same as
Fig.1, but showing the cross section of $H A$ associated production
at the LHC versus $m_H$.}
\end{figure}

Furthermore, we examine other complementary new channels for
detecting the Higgs sector of the L2HDM. Firstly, we check the
associated $H A$ production at the LHC. The cross section of this
process is shown in Fig. 4 and one can learn that for $m_H < 140$
GeV the cross section is larger than 100 fb. The dominant decay of
$H$ in this case is found to be $H\to A A$, with a branching ratio
larger than $80\%$,  so the main signal of this process is $6 \tau$.
Due to the lightness of $A$, great efforts are needed to
analyze the signal and the backgrounds in order to detect
this process at the LHC.
Secondly, we note that $A$ is always lighter than $Z$ boson in
the allowed parameter space and thus it may be produced from $Z$
decays. So we investigate the decay $Z \to \tau \bar{\tau} A$ and
find its branching ratio ranging from $10^{-5}$ to $10^{-4}$ for
$m_A < 40$ GeV (corresponding to $98\%$ of the allowed parameter
space). Such a large rate is within the sensitivity of the GigaZ
option at the proposed International Linear Collider \cite{ILC}.

\section{Conclusion}
\label{sec5}
We examined various direct and indirect constraints on
the lepton-specific two-Higgs doublet model and then checked the
property of the Higgs bosons in the allowed parameter space. We
found that the allowed space has a very light CP-odd Higgs boson $A$
($m_A < 30$ GeV with $95\%$ possibility) which is composed
dominantly by the leptonic Higgs and decays dominantly into $\tau^+
\tau^-$. The SM-like Higgs boson $h$ decays dominantly in the
mode $h \to A A \to 4 \tau $, which may make the Higgs discovery
difficult at the LHC. We also checked other possibilities for
testing the Higgs sector of this model and found that the decay $Z
\to \tau^+ \tau^- A$ has a branching ratio ranging from $10^{-5}$ to
$ 10^{-4}$, which may be accessible at the the GigaZ option of the
ILC.

\section*{Acknowledgement}
This work was supported in part by the National Natural Science
Foundation of China (NNSFC) under grant Nos. 10505007, 10821504,
10725526 and 10635030, by HASTIT under grant No. 2009HASTIT004, by
the Project of Knowledge Innovation Program (PKIP) of Chinese
Academy of Sciences under grant No. KJCX2.YW.W10 and by an
invitation fellowship of LHC Physics Focus Group, National Center
for Theoretical Sciences, Taiwan.

\end{document}